\documentclass[twocolumn,prd,showpacs,showkeys,nofootinbib]{revtex4-1}
\usepackage{enumerate}
\usepackage{amsmath,amssymb,pgf,graphicx,textpos,color}
\usepackage{url}

\definecolor{mygray}{gray}{0.5}
\begin{document}

\title{Constraints on the diffuse photon flux with energies above
$10^{18}$~eV using the surface detector of the Telescope~Array
experiment}

\author{R.U.~Abbasi$^{1}$,
M.~Abe$^{2}$,
T.~Abu-Zayyad$^{1}$,
M.~Allen$^{1}$,
R.~Azuma$^{3}$,
E.~Barcikowski$^{1}$,
J.W.~Belz$^{1}$,
D.R.~Bergman$^{1}$,
S.A.~Blake$^{1}$,
R.~Cady$^{1}$
% ...
\\~\\
{\footnotesize\it
$^{1}$ High Energy Astrophysics Institute and Department of Physics and Astronomy, University of Utah, Salt Lake City, Utah, USA \\
$^{2}$ The Graduate School of Science and Engineering, Saitama University, Saitama, Saitama, Japan \\
$^{3}$ Graduate School of Science and Engineering, Tokyo Institute of Technology, Meguro, Tokyo, Japan \\
$^{4}$ Department of Physics and The Research Institute of Natural Science, Hanyang University, Seongdong-gu, Seoul, Korea \\
$^{5}$ Department of Physics, Tokyo University of Science, Noda, Chiba, Japan \\
$^{6}$ Department of Physics, Kindai University, Higashi Osaka, Osaka, Japan \\
$^{7}$ Service de Physique ThÃ©orique, UniversitÃ© Libre de Bruxelles, Brussels, Belgium \\
$^{8}$ Institute for Cosmic Ray Research, University of Tokyo, Kashiwa, Chiba, Japan \\
$^{9}$ Graduate School of Science, Osaka City University, Osaka, Osaka, Japan \\
$^{10}$ Kavli Institute for the Physics and Mathematics of the Universe (WPI), Todai Institutes for Advanced Study, University of Tokyo, Kashiwa, Chiba, Japan \\
$^{11}$ Information Engineering Graduate School of Science and Technology, Shinshu University, Nagano, Nagano, Japan \\
$^{12}$ Faculty of Engineering, Kanagawa University, Yokohama, Kanagawa, Japan \\
$^{13}$ Interdisciplinary Graduate School of Medicine and Engineering, University of Yamanashi, Kofu, Yamanashi, Japan \\
$^{14}$ Astrophysical Big Bang Laboratory, RIKEN, Wako, Saitama, Japan \\
$^{15}$ Department of Physics, Sungkyunkwan University, Jang-an-gu, Suwon, Korea \\
$^{16}$ Department of Physics, Tokyo City University, Setagaya-ku, Tokyo, Japan \\
$^{17}$ Institute for Nuclear Research of the Russian Academy of Sciences, Moscow, Russia \\
$^{18}$ Advanced Research Institute for Science and Engineering, Waseda University, Shinjuku-ku, Tokyo, Japan \\
$^{19}$ Department of Physics, Chiba University, Chiba, Chiba, Japan \\
$^{20}$ Department of Physics, School of Natural Sciences, Ulsan National Institute of Science and Technology, UNIST-gil, Ulsan, Korea \\
$^{21}$ Department of Physics, Yonsei University, Seodaemun-gu, Seoul, Korea \\
$^{22}$ Academic Assembly School of Science and Technology Institute of Engineering, Shinshu University, Nagano, Nagano, Japan \\
$^{23}$ Faculty of Science, Kochi University, Kochi, Kochi, Japan \\
$^{24}$ Department of Physical Sciences, Ritsumeikan University, Kusatsu, Shiga, Japan \\
$^{25}$ Sternberg Astronomical Institute, Moscow M.V. Lomonosov State University, Moscow, Russia \\
$^{26}$ Department of Physics and Astronomy, Rutgers University - The State University of New Jersey, Piscataway, New Jersey, USA \\
$^{27}$ Earthquake Research Institute, University of Tokyo, Bunkyo-ku, Tokyo, Japan \\
$^{28}$ Department of Engineering Science, Faculty of Engineering, Osaka Electro-Communication University, Neyagawa-shi, Osaka, Japan \\
$^{29}$ Graduate School of Information Sciences, Hiroshima City University, Hiroshima, Hiroshima, Japan \\
$^{30}$ Institute of Particle and Nuclear Studies, KEK, Tsukuba, Ibaraki, Japan \\
$^{31}$ National Institute of Radiological Science, Chiba, Chiba, Japan \\
$^{32}$ CEICO, Institute of Physics, Czech Academy of Sciences, Prague, Czech Republic \\
$^{33}$ Department of Physics and Institute for the Early Universe, Ewha Womans University, Seodaaemun-gu, Seoul, Korea \\
$^{34}$ Department of Physics, Ehime University, Matsuyama, Ehime, Japan\\
$^{35}$ Faculty of Physics, M.V. Lomonosov Moscow State University, Moscow, Russia
}}\let\thefootnote\relax\footnote{$^{*}$ Deceased}
\let\thefootnote\relax\footnote{$^{**}$ Corresponding author, grisha@ms2.inr.ac.ru}

\begin{abstract}
We present the results of the search for ultra-high-energy photons
with nine years of data from the Telescope Array surface detector. A
multivariate classifier is built upon 16 reconstructed parameters of
the extensive air shower. These parameters are related to the
curvature and the width of the shower front, the steepness of the
lateral distribution function, and the timing parameters of the
waveforms sensitive to the shower muon content. A total number of two
photon candidates found in the search is fully compatible with the
expected background. The $95\%$\,CL limits on the diffuse flux of the
photons with energies greater than $10^{18.0}$, $10^{18.5}$,
$10^{19.0}$, $10^{19.5}$ and $10^{20.0}$~eV are set at the level of
$0.067$, $0.012$, $0.0036$, $0.0013$,
$0.0013~\mbox{km}^{-2}\mbox{yr}^{-1}\mbox{sr{}}^{-1}$ correspondingly.
\end{abstract}
\maketitle

\section{Introduction}

The Telescope Array~(TA) experiment~~\cite{TASD,TAFD} is a hybrid detector
operating in Utah, USA. TA consists of a surface detector array of 507
plastic scintillators with 1.2~km spacing covering approximately 700~km$^2$ area and
three fluorescence detectors. The purpose of this {\it Paper} is to
present the limits on diffuse photon flux with energies
greater than $10^{18}$~eV based on nine years of surface detector
operation. 

Several limits on ultra-high-energy diffuse photon flux have been set
by independent experiments, including Haverah Park, AGASA, Yakutsk,
Pierre Auger and TA observatories
~\cite{HP_lim,AGASA_1stlim,Ylim,Ylim18,AGASA_Risse,A+Y,Auger_fdlim,Auger_sdlim1,Auger_sdlim2,Auger_hyblim,TAglim},
but no evidence for primary photons has been found at present. The
upper limit on a photon flux from Southern Hemisphere point sources is
set by the Pierre Auger Observatory~\cite{Auger_point}. Photon flux
limits may be used to constrain the parameters of top-down
models~\cite{Berezinsky:1998ft} as well as the properties of
astrophysical sources and their evolution in the scenario of
Greisen-Zatsepin-Kuzmin~\cite{g,zk} cut-off. The flux of
ultra-high-energy photons is the most pronounced signature for the
heavy decaying dark matter
searches~\cite{Kalashev:2016cre,Kalashev:2017ijd}.  Moreover, the
results of the photon search severely constrain the parameters of
Lorentz invariance violation at Planck
scale~\cite{Coleman:1998ti,Galaverni:2007tq,Maccione:2010sv,Rubtsov:2012kb,Rubtsov:2013wwa}.
Finally, photons with energies above $\sim 10^{18}$~eV might be
responsible for CR events correlated with BL Lac type objects on the
angular scale significantly smaller than the expected deflection of
protons in cosmic magnetic fields and thus suggesting neutral
primaries~\cite{Gorbunov:2004bs,Abbasi:2005qy} (see Ref.~\cite{axion}
for a particular mechanism).

\section{Data set and simulations}

We use the TA surface detector data set covering nine years
of observation from May 11, 2008 to May 10, 2017. The surface detector (SD) has
been collecting data for more than 95\% of time during that
period~\cite{TASDspec,TASDspecICRC}.

Air showers induced by primary photons differ significantly from the
hadron-induced events (see e.g.~\cite{RisseRev} for a review). The
TA SD stations contain two layers of 1.2 cm
think plastic scintillators with an area of 3~m$^2$ which detect both muon and
electromagnetic components of the extensive air shower and therefore
are sensitive to showers induced by primary photons~(see
Ref.~\cite{Nih} for discussion).

We employ Monte-Carlo simulations of TA SD events induced by simulated
proton and photon-induced extensive air showers. The proton induced
simulated event set is used as a background while the photon set is
used as a signal. We produce simulated events by
CORSIKA~\cite{corsika} with EGS4~\cite{Nelson:1985ec} model for
electromagnetic interactions, PRESHOWER code~\cite{Homola:2003ru} for
interactions of photons in geomagnetic field, QGSJET~II-03
\cite{QGSJET-II} and FLUKA~\cite{fluka} for high and low energy
hadronic interactions. The showers are simulated with thinning
($\epsilon=10^{-6}$) and the dethinning procedure is used to recover
small scale structure of the shower
fluctuations~\cite{dethinning}. The proton Monte-Carlo set is
simulated with the primary energy spectrum of HiRes
experiment~\cite{HiRes_spec} and is validated by a direct comparison
with the TA SD data~\cite{StokesMC}. The classification method of the
present Paper requires high Monte-Carlo statistics at each energy
range. Therefore, an additional high-energy proton Monte-Carlo set is
simulated with a starting energy of $10^{18.5}$\,eV, which is used for
energy ranges above $10^{19.5}$\,eV. The photon set is simulated
following $E^{-1}$ power spectrum and then sampled in each energy
range according to $E^{-2}$ spectrum for compatibility with the
assumptions of the photon searches performed by other groups.

\begin{figure*}[t]
\begin{center}
  \includegraphics[width=0.90\columnwidth]{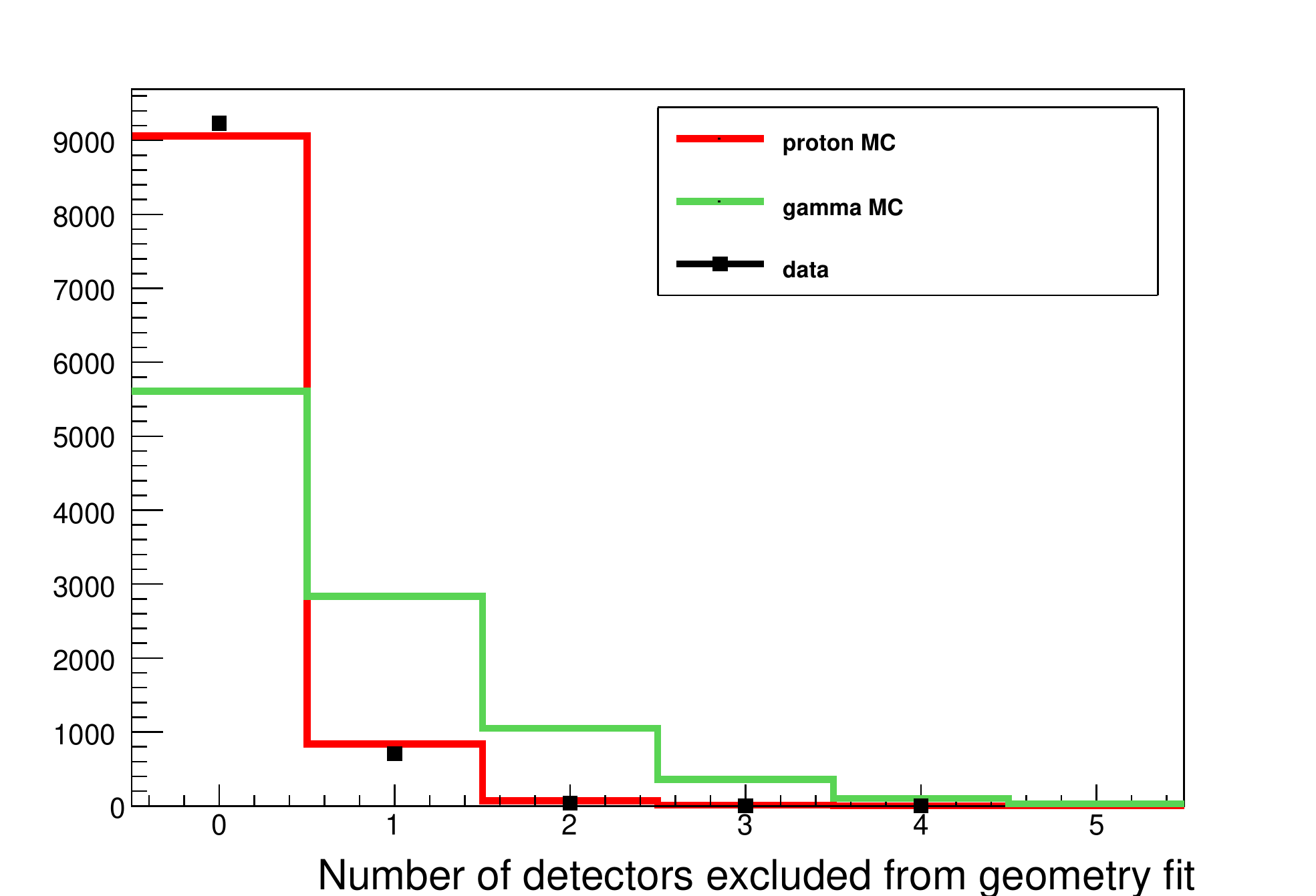}
  \includegraphics[width=0.90\columnwidth]{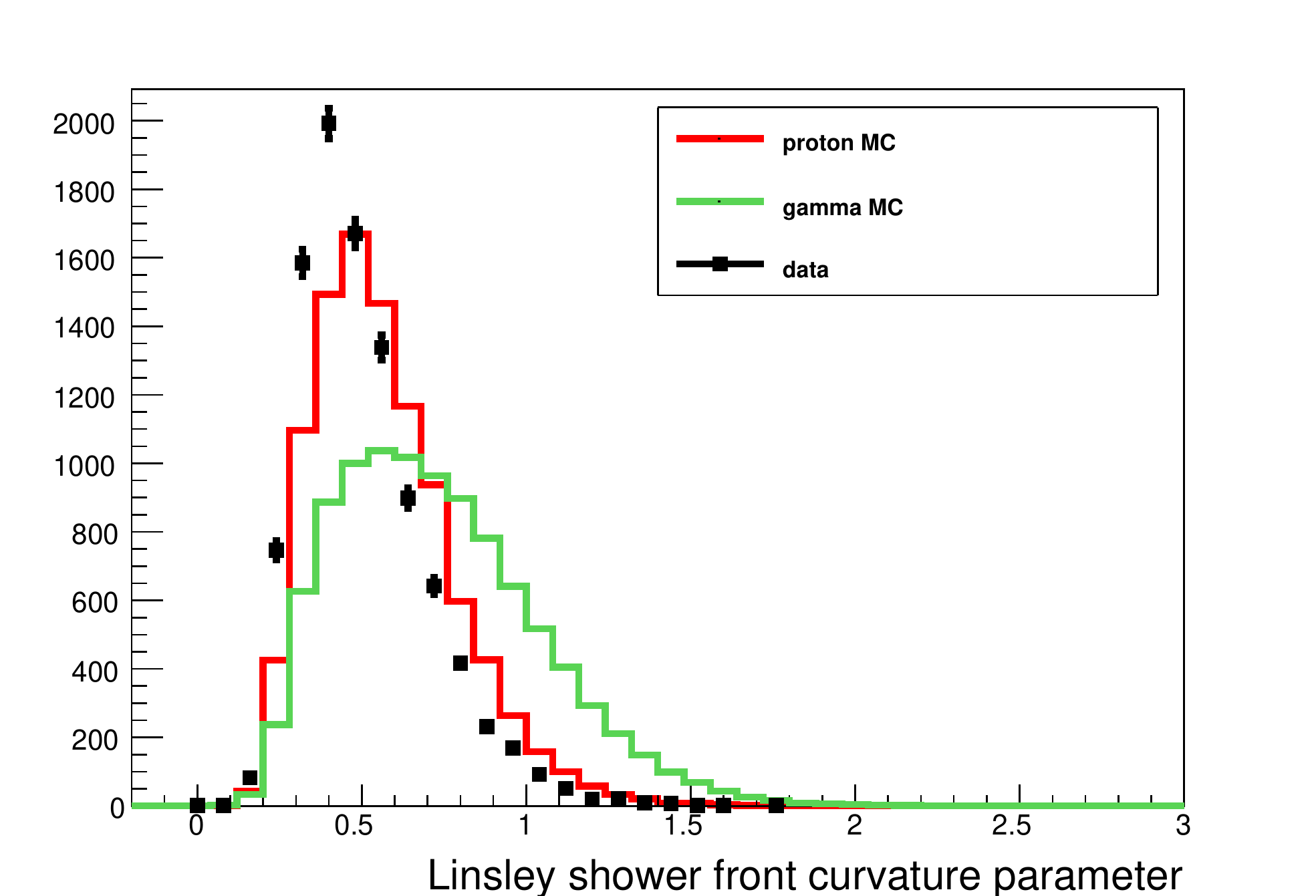}
\caption{\label{a_ngex_dist} Distributions of the two observables with
  the strongest $\gamma$-p separation power for data
  (black) compared with proton and photon-induced Monte-Carlo events
  for $E>10^{19}$~eV
  (solid red - protons, dashed green - photons).  }
\end{center}
\end{figure*}

Detector response is accounted for by using look-up tables simulated
with GEANT4~\cite{GEANT}.  Real-time array configuration and detector
calibration information of the nine years of TA SD observations are
used for each simulated event. Monte-Carlo~(MC) events are produced in
the same format as real events and the analysis procedures are applied
in the same way to both~\cite{StokesMC}. The statistics of the proton
MC set is 5.5 times larger than the number of the observed events.

\section{Reconstruction and observables}
\label{sec_Reconstruct}

\begin{figure*}[t]
\begin{center}
  \includegraphics[width=0.9\columnwidth]{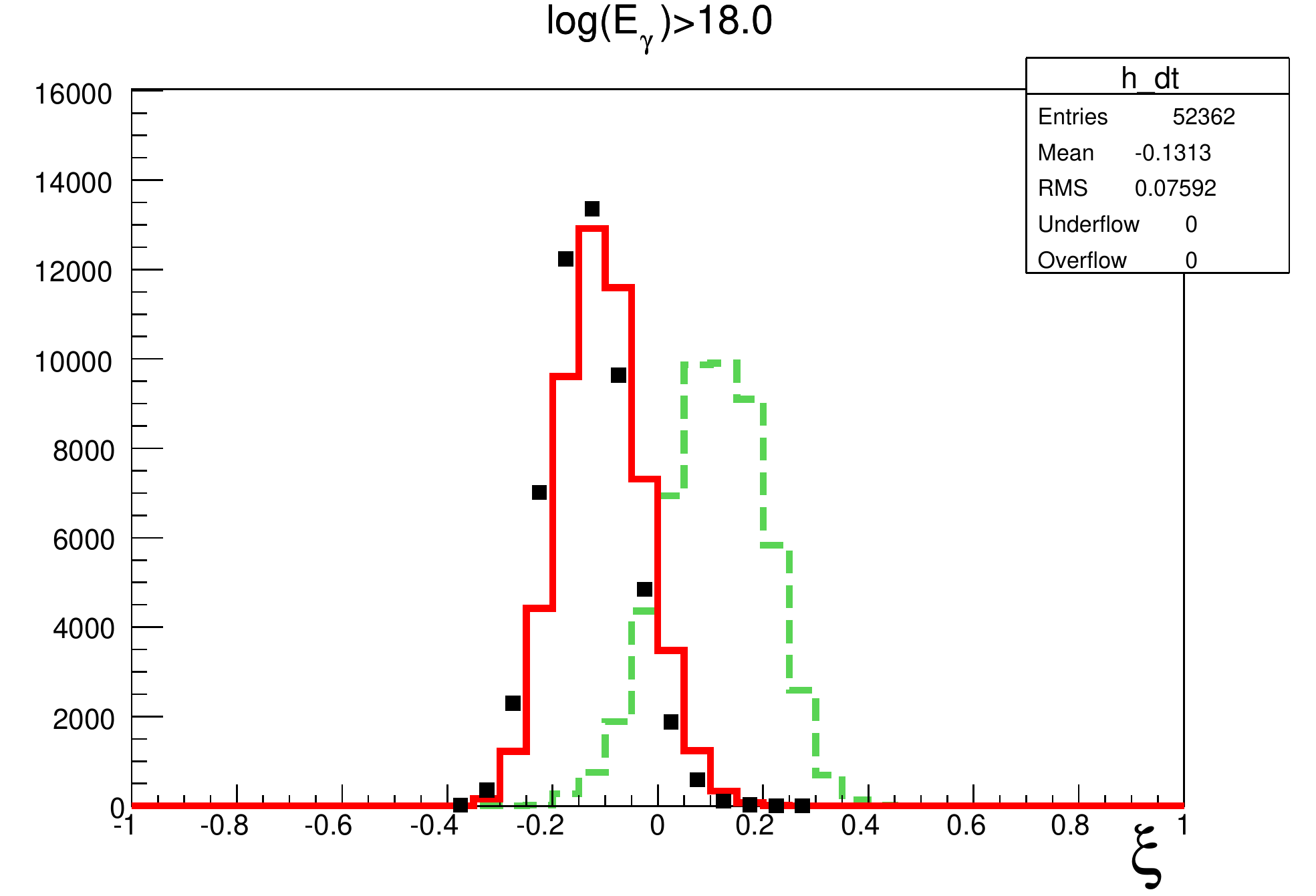}
  \includegraphics[width=0.9
\columnwidth]{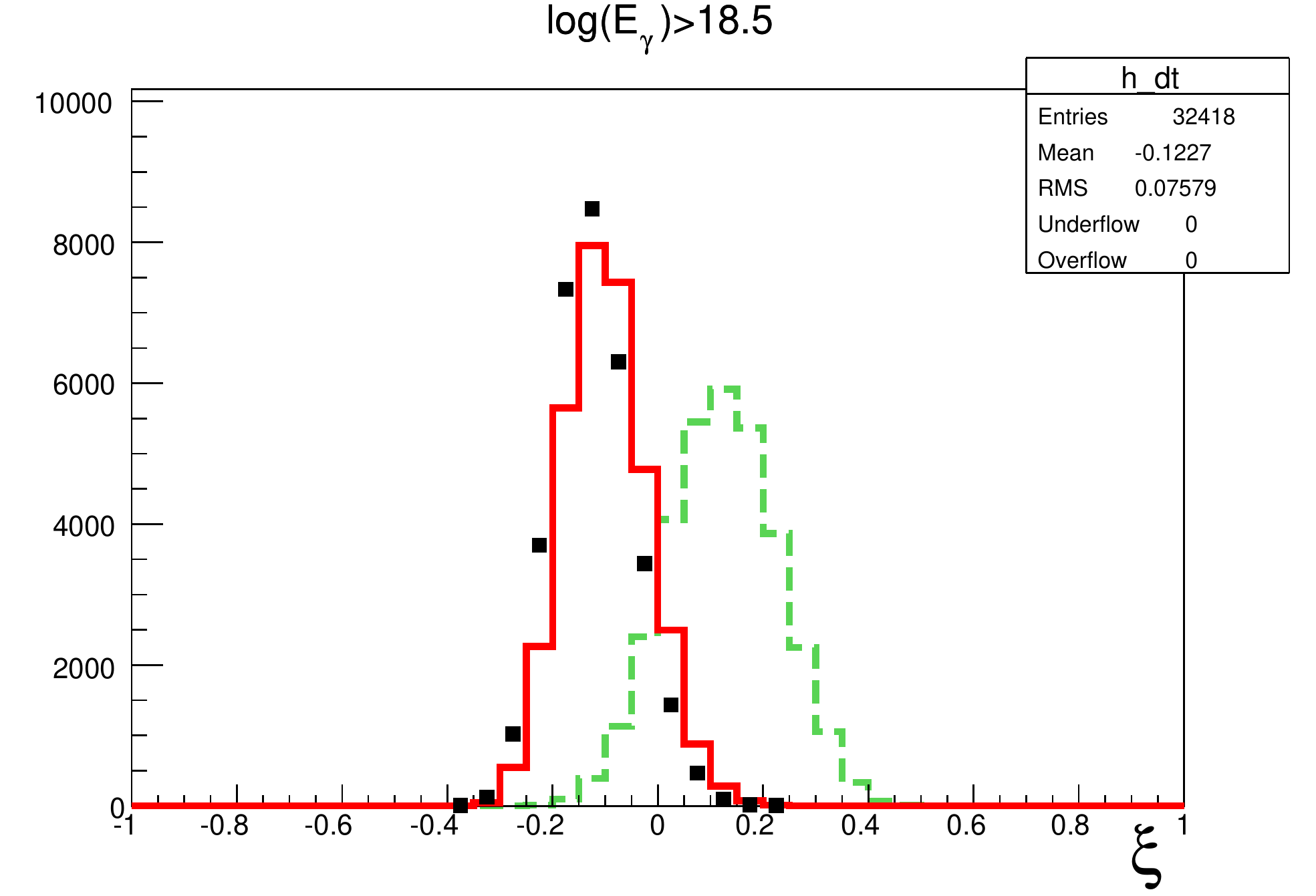}\\
  \includegraphics[width=0.9\columnwidth]{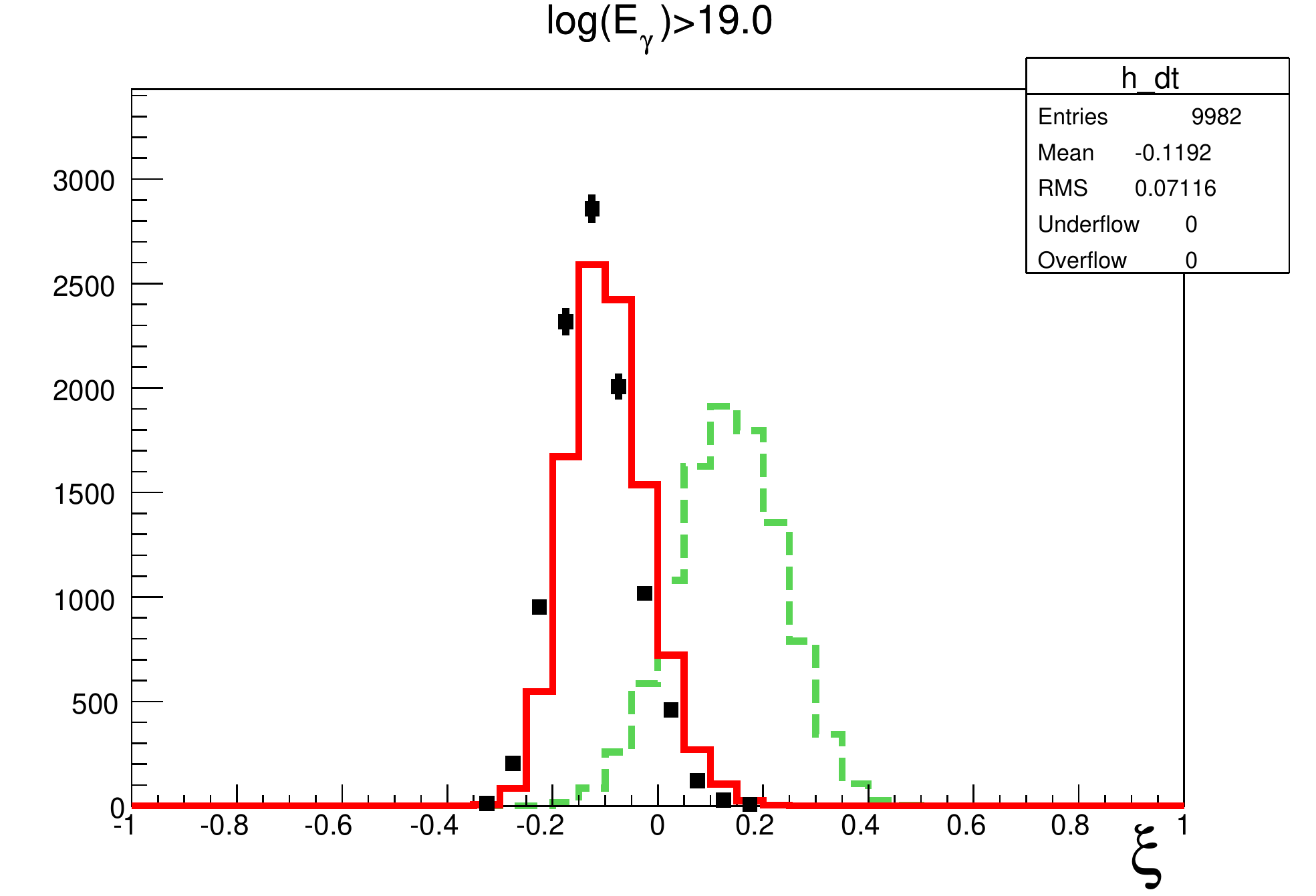}
  \includegraphics[width=0.9\columnwidth]{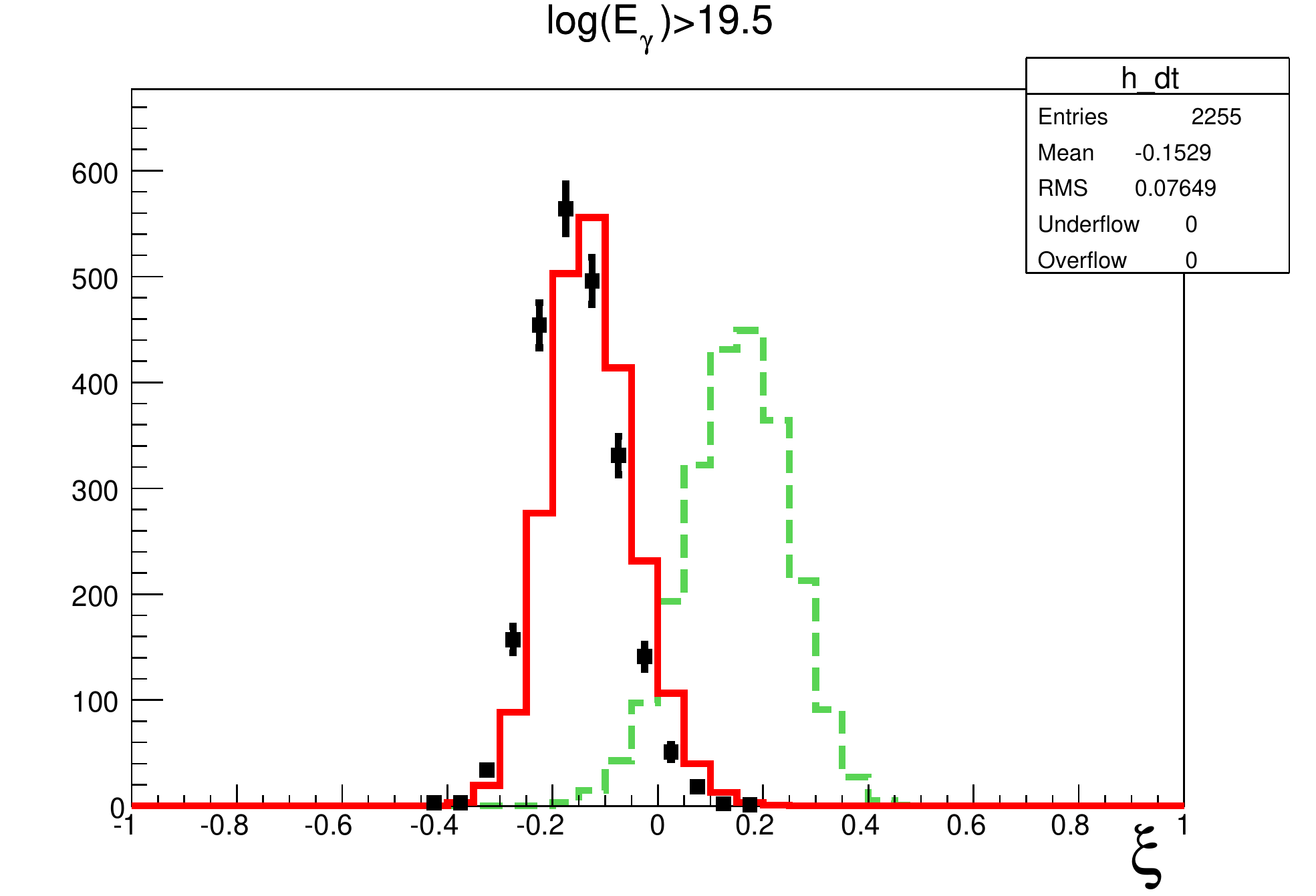}\\
  \includegraphics[width=0.9\columnwidth]{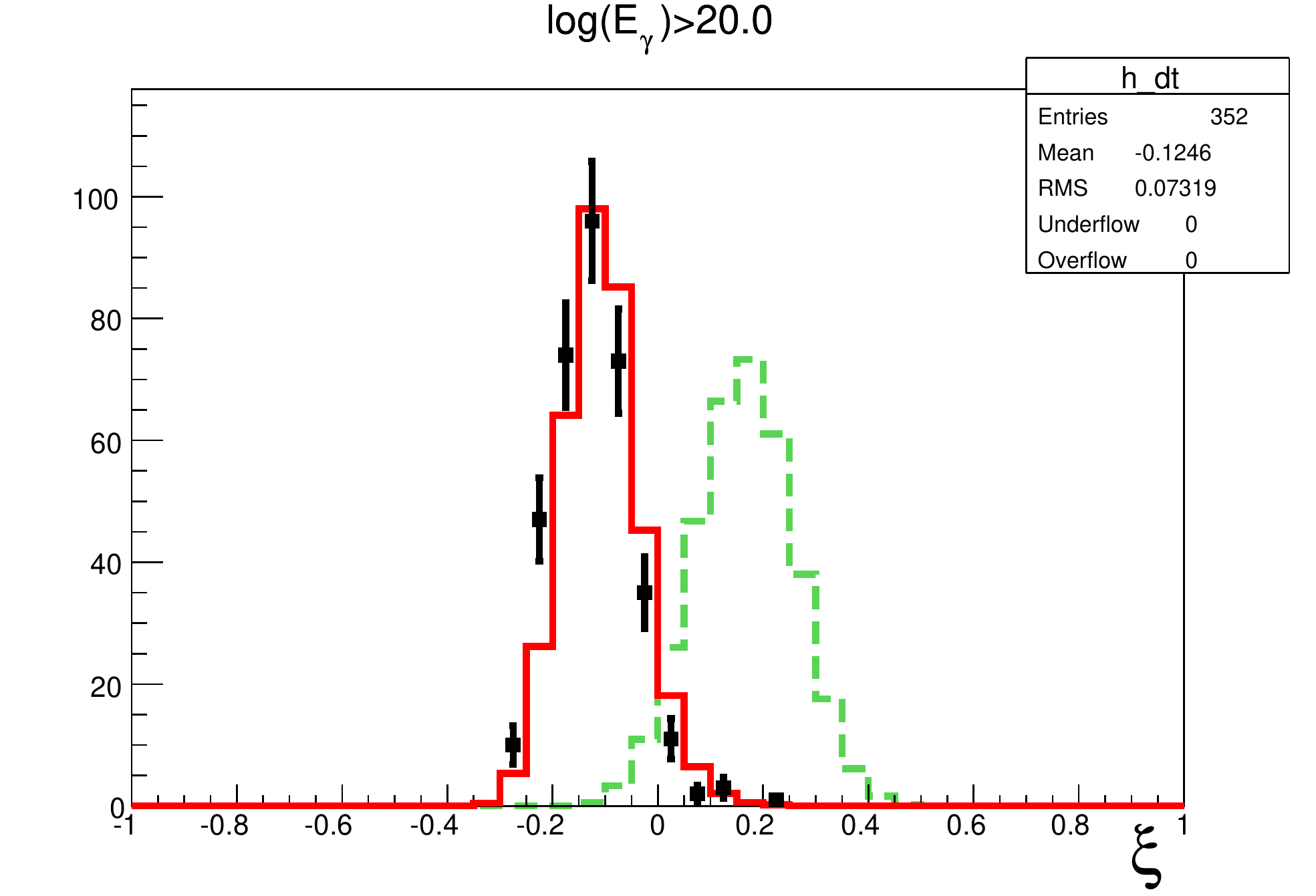}
\caption{\label{xi_dist} Distributions of the $\xi$ parameter for data
  (black) compared with proton and photon-induced Monte-Carlo events
  for the five energy ranges of interest
  (solid red - protons, dashed green - photons).  }
\end{center}
\end{figure*}

We reconstruct each event with a joint fit of the geometry and lateral
distribution function (LDF) and determine Linsley shower front
curvature parameter ``$a$''~\cite{Linsley:1975hr} along
with the arrival direction, core location and signal density at 800
meters $\mathcal S \equiv S_{800}$. The same reconstruction procedure
is applied to both data and MC events.

Reconstruction of each data and MC event results in a set of 16
observables, which are described in~\cite{TASDcompos}. The following
observables are used for construction of a multivariate
classification method:

\begin{enumerate}
\item Zenith angle, $\theta$; 
\item Signal density at 800\ m from the shower core, $S_{800}$;  
\item Linsley front curvature parameter, $a$ obtained front the fit of
  the shower front with the
  AGASA-modified Linsley time delay function~\cite{Teshima:1986rq};
\item Area-over-peak (AoP) of the signal at 1200 m~\cite{Auger_nuAOP};
\item AoP slope parameter~\cite{ACAT2014};
\item Number of stations with Level-1 trigger~\cite{TASD};
\item Number of stations excluded from the fit of the shower front
  due to large contribution to $\chi^2$;
\item $\chi^2/d.o.f.$;
\item $S_b$ parameter for $b=3$; $S_b$ is defined
as $b$-th moment of the LDF:
\begin{equation}
S_b = \sum\limits_{i} \left[S_i \times \left(r_i/r_0\right)^b\right]\,,
\end{equation}
where $S_i$ is the signal of $i$-th station, $r_i$ is the distance
from the shower core to a given station, $r_0=1000$\,m. The sum is
calculated over all triggered non-saturated stations. The $S_b$ is proposed as a composition-sensitive parameter at~\cite{MedinaSb};
\item $S_b$ parameter for $b=4.5$;
\item The sum of signals of all stations of the event;
\item An average asymmetry of signal at upper and lower layers of the
  stations defined as:
\begin{equation}
\mathcal{A} = \frac{\sum\limits_{i,\alpha}
  |s^{upper}_{i,\alpha}-s^{lower}_{i,\alpha}|}{\sum\limits_{i,\alpha}
  |s^{upper}_{i,\alpha}+s^{lower}_{i,\alpha}|}\,,
\end{equation}
where $s^{upper|lower}_{i,\alpha}$ is the FADC value of upper or lower
layer of $i$-th station at $\alpha$-th time bin. The sum is calculated
over all triggered non-saturated stations over all time bins of the
corresponding FADC traces;
\item Total number of peaks of FADC trace summed over both upper and
  lower layers of all stations participating in the event. To suppress
  accidental peaks as a result of FADC noise, we define a peak as a
  time bin with a signal above 0.2 Vertical equivalent muons (VEM)
  which is higher than a signal of the 3 preceding and 3 consequent
  time bins;
\item Number of peaks for the detector with the largest signal;
\item Total number of peaks present in the upper layer and not in lower;
\item Total number of peaks present in the lower layer and not in upper.
\end{enumerate}

For each real event {\it ``i''} we estimate the energy of a hypothetical
photon primary $E^i_\gamma = E_\gamma(\mathcal S^i, \theta^i, \phi^i)$,
i.e. the average energy of the primary photon, inducing a shower
with the same arrival direction and $\mathcal S$. A look-up table for
$E_\gamma(\mathcal S, \theta, \phi)$ is built using photon MC set.

Both data and MC events are selected by the following quality criteria:
\begin{enumerate}[(a)]
\item Zenith angle: $0^\circ < \theta < 60^\circ$;
\item The number of stations triggered is 7 or more;
\item Shower core is inside the array boundary with the distance to
  the boundary larger than 1200 meters;
\item Joint LDF and shower front fit quality, $\chi^2/$d.o.f.$<5$;
\item $E_\gamma(S^i_{800}, \theta^i, \phi^i) > E_0$~eV for photon
  search, where $E_0=10^{18.0},10^{18.5},10^{19.0},10^{19.5}$, or
  $10^{20.0}$~eV is the lower limit of the energy range;
\item Event is not time-correlated with the lightnings in the Vaisala
  lightning database~\cite{NLDN1,NLDN2,NLDNurl}. The details of the cut are given below.
\end{enumerate}

\begin{figure*}[t]
\begin{center}
  \includegraphics[width=0.9\columnwidth]{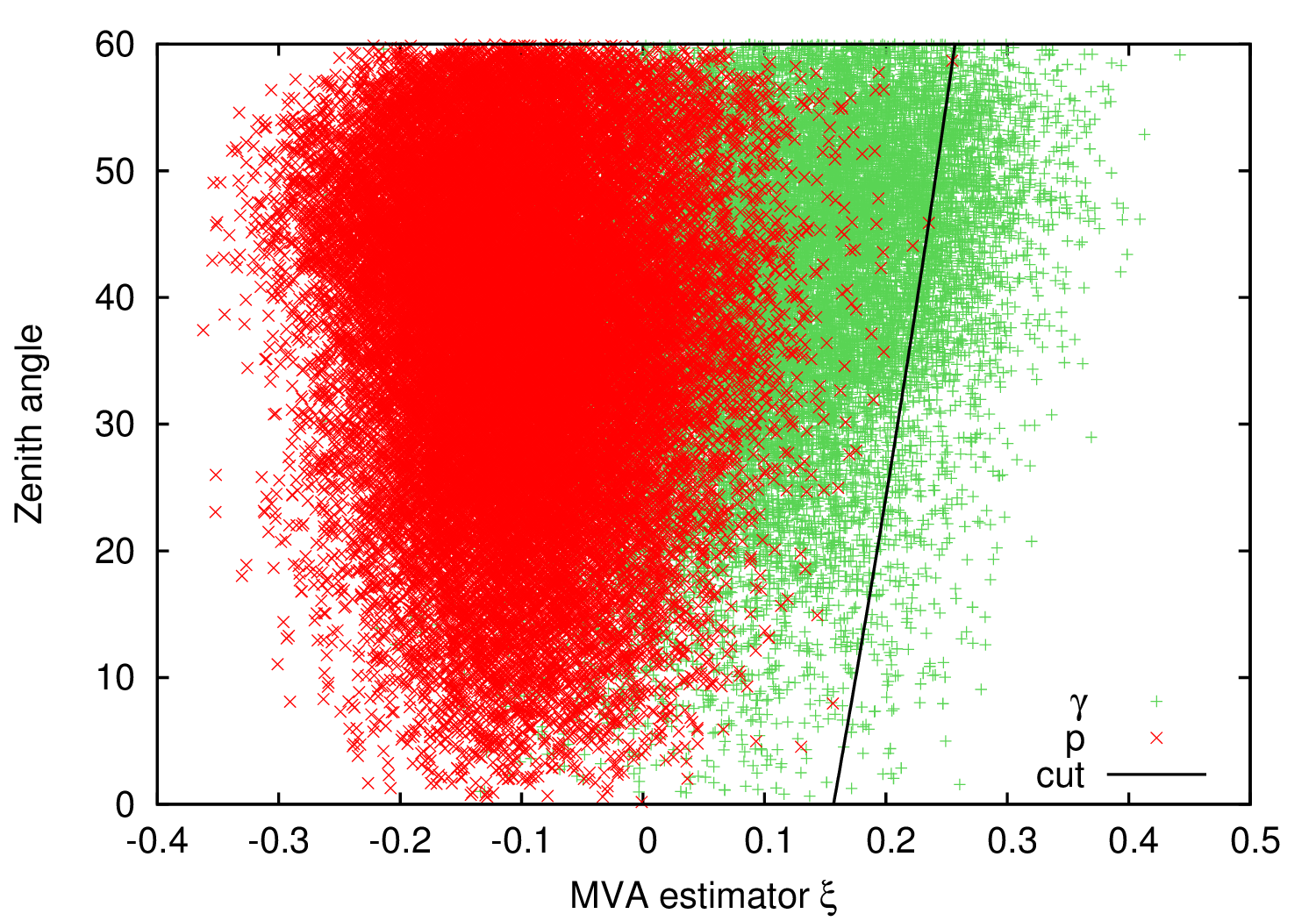}
  \includegraphics[width=0.9\columnwidth]{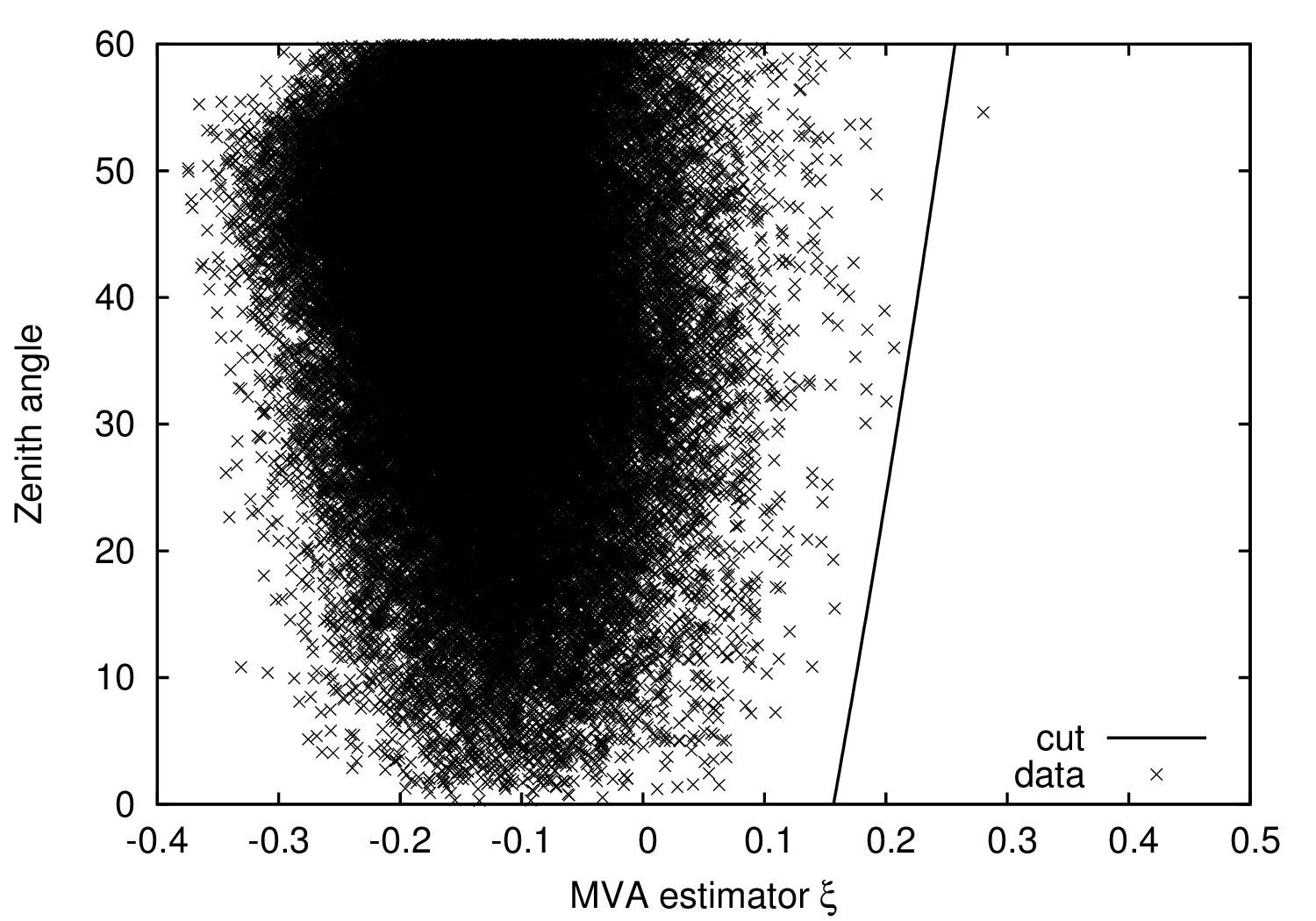}
\caption{\label{mva_cut} Left panel: The scatter-plot of $\xi$ and
  zenith angle for proton (red dots) and photon (green dots)
  Monte-Carlo sets along with the optimal $\xi$-cut (black line) for
  the energy range $E>10^{18}$~eV. Right panel: the same $\xi$-cut
  applied to the data set.}
\end{center}
\end{figure*}

The conditions (a)-(e) are the same as the quality cuts in the previous TA
analysis~\cite{TAglim}, while the condition (f) is introduced in the present
analysis. The details of the latter are provided below.

It was shown that some of the extensive air showers observed by the TA
SD are produced by the terrestrial gamma-ray flashes
(TGFs)~\cite{Okuda,TALMA}. Being initiated by the gamma-rays in the
middle of the atmosphere, these cascades share many properties of the
showers induced by the ultra-high-energy photons. Namely, the showers
induced by TGFs contain no muons and possess large curvature of the
front. In order to exclude possible photon candidates of the
atmospheric origin we use the Vaisala lightning database from the
U.S. National Lightning Detection Network
(NLDN)~\cite{NLDN1,NLDN2,NLDNurl}. We obtained the list of the NLDN
lightning events located within a 15 miles radius of the Central Laser
Facility of the TA in the time range of the study 2008-05-11 --
2017-05-10. The number of the events in the list is 31622 and they are
grouped in time in such a way that a total of 910 astronomical hours
contain one or more lightning events. With condition (f), we remove
the SD events which occur within 10 minutes time interval before or
after the NLDN lightnings. The cut efficiently removes all the events
known to be related to the TGFs with the cost of only $0.66\%$ of the
exposure time.

After the cuts, the data set includes 52362 events. The number of
events in proton and high-energy proton Monte-Carlo sets is 283k and
662k, correspondingly. The photon Monte-Carlo set includes 57k, 151k,
325k, 354k and 330k reconstructed events for the energy ranges defined
with $E_0=10^{18.0},10^{18.5},10^{19.0},10^{19.5}$, and $10^{20.0}$
correspondingly.

\subsection{Method}

The analysis is based on a proton-photon classification procedure
using the method of boosted decision trees (BDT). Following the
analyses of~\cite{Auger_point,ACAT2014} we use the {\it TMVA}
package~\cite{TMVA} for {\sc root}~\cite{ROOT} as an implementation of the
method. The decision forest is constructed using the 16 observable
parameters listed in the Section. The BDT is
trained using the proton Monte-Carlo set as a background and the
photon Monte-Carlo as a signal. The Monte-Carlo set is split into
three subsets of equal number of events: (I) for training the classifier, (II) for cut
optimization, (III) for exposure estimate. For the photon search, the
classifier is built independently for each energy range of interest
$E>E_0$, where $E_0$ takes values of $10^{18.0}$, $10^{18.5},
10^{19.0}, 10^{19.5}$ and $10^{20.0}$\,eV.

The distributions of the 16 parameters in the SD data are in a
reasonable agreement with the proton MC, see individual parameter
distributions in~\cite{TASDcompos}. The {\it TMVA} classifier ranks
the variables according to importance parameter, which indicates
relative contribution of each parameter to separation power. In the
present study, all the parameters show an importance value between 4\%
and 9\% with the strongest separation power for the number of
detectors excluded from the geometry fit (9\%) and Linsley shower
front (7\%) curvature. The histograms of these two parameters for data
and simulated events for the energy range $E>10^{19}$~eV are shown
in Figure~\ref{a_ngex_dist}. The result of the BDT classifier is a
single parameter $\xi^i$ for each event {\it ``i''} which by
definition has a bounded range $-1 \le \xi^i \le 1$. The
$\xi$-parameter is finally used for a one-parametric composition
analysis. The histograms of the $\xi$-parameter for data and simulated
events for the five energy ranges of interest are shown in
Figure~\ref{xi_dist}. The histograms indicate general compatibility of
the data with the proton Monte-Carlo, while a shift to the smaller
$\xi$ values may be observed. The latter corresponds to either
moderately heavier hadronic primaries or hadronic model uncertainty,
see discussion of the latter in~\cite{TASDcompos}.

The photon candidates are selected with the zenith angle dependent cut
on~$\xi$ 
\begin{equation}
\xi > \xi_{cut}(\theta)\,.
\end{equation}
The cut function is approximated as a quadratic polynomial of
$\theta$. The cut is optimized with the part II of the MC with the merit
factor defined as an average photon flux upper limit in the case of
null-hypothesis that all events in the data set are protons. The
$\xi$-cut for $E>10^{18}$~eV is shown in Figure~\ref{mva_cut} along
with the Monte-Carlo (left panel) and data (right panel).

\begin{figure}[t]
\begin{center}
  \includegraphics[height=1.0\columnwidth,angle=270]{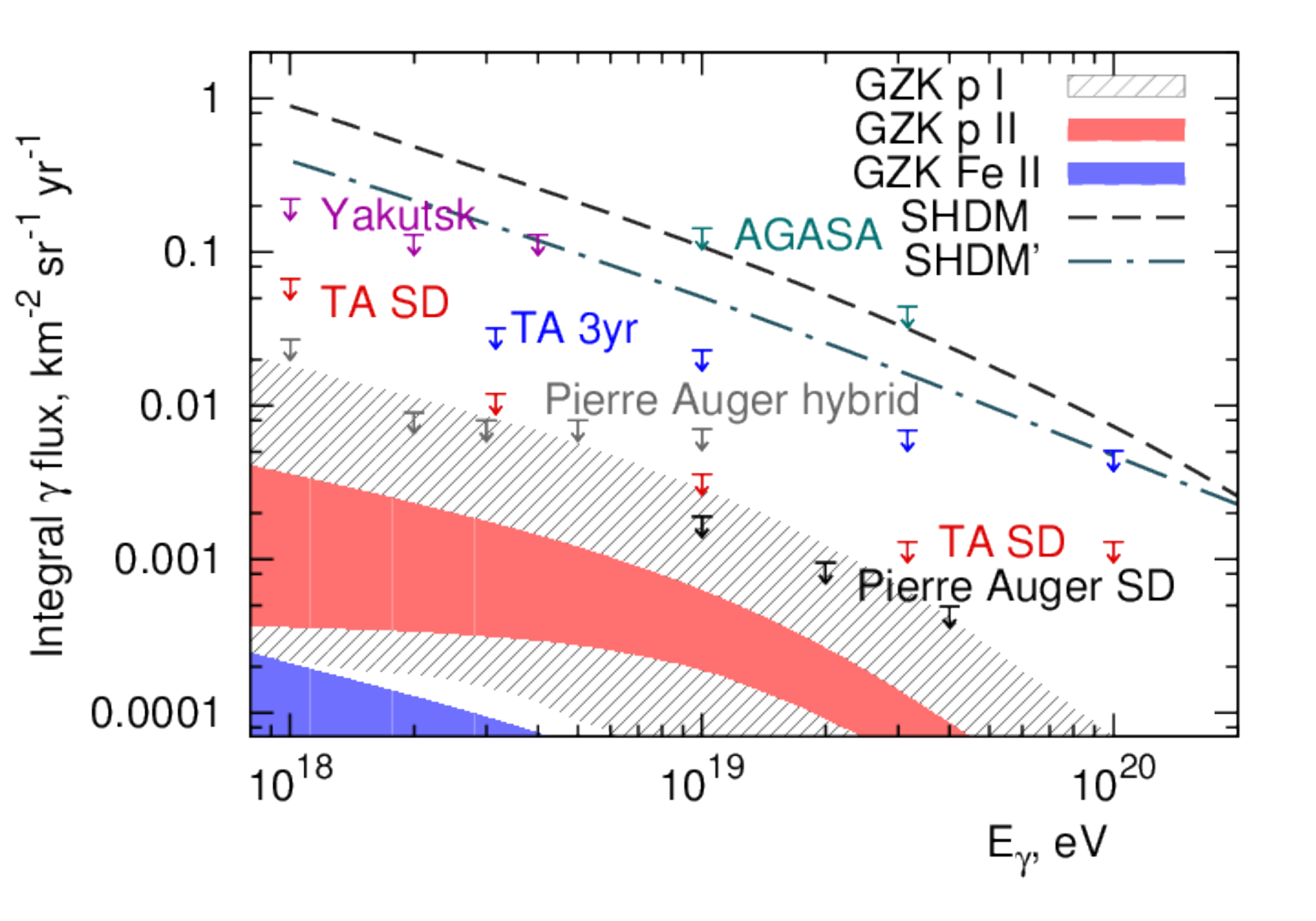} 
\end{center}
\caption{\label{fig:glresult}
The photon flux limit presented in this {\it Paper}~(TA SD, red arrows) compared with the
results from AGASA~(light blue)~\cite{AGASA_1stlim}, Pierre Auger Observatory
SD~(black)~\cite{Auger_sdlim1,Auger_sdlim2} and hybrid data~(gray)~\cite{Auger_hyblim}, 
Yakutsk~(magenta)~\cite{Ylim18} and previously published TA SD result
(TA 3yr, dark blue)~\cite{TAglim} and
the predictions of some models~\cite{Gelmini:2005wu,Hooper:2010ze,MMWG2014}.
}
\end{figure}

\section{Photon search results}
\label{sec_Results}
The geometrical exposure for the considered SD observation period with $0^\circ <
\theta < 60^\circ$ and the boundary cut is given by
\begin{equation}
A_{geom} = 12060~\mbox{km}^{2}~\mbox{sr~yr}\,.
\end{equation}

After the cuts, the effective exposures $A_{eff}^\gamma$ for different
$E_0$ values are given in
Table~\ref{tab:exposure}.

\begin{table}[h]
\begin{center}
\begin{tabular}{|c|c|c|c|}
\hline
$E_0$ & quality cuts (a)-(f) & $\xi$-cut & $A_{eff}^\gamma$,\,$\mbox{km}^{2}\,\mbox{yr}\,\mbox{sr}$ \\
\hline
$10^{18.0}$ & 6.5\% & 9.8\% & {\bf 77} \\ \hline
$10^{18.5}$ & 19.9\% & 10.6\% & {\bf 255} \\ \hline
$10^{19.0}$ & 43.6\% & 16.2\% & {\bf 852} \\ \hline
$10^{19.5}$ & 52.0\% & 37.2\% & {\bf 2351} \\ \hline
$10^{20.0}$ & 64.2\% & 52.3\% & {\bf 4055} \\ \hline
\end{tabular}
\end{center}
\caption{\label{tab:exposure} Contribution of the cuts to an effective
  exposure in the energy ranges of interest. The value represents the ratio
  of the exposure after the given cut to the exposure before cut.}
\end{table}

No photon candidates are found in the data set
for $E_0=10^{18.5}$, $10^{19.0}$ and $10^{19.5}$~eV. There is one
candidate for each of the two energy ranges $E_0=10^{18.0}$ and
$10^{20.0}$\,eV.  An upper limit on a mathematical expectation of
the number of photons is determined following
Ref.~\cite{Feldman:1997qc}. The flux upper limits are given by the
relation
\begin{equation}
\bar n_{\gamma} = F_{\gamma} A_{eff}^\gamma\,.
\end{equation}
The resulting 95\% CL photon diffuse flux upper limits are summarized
in Table~\ref{tab:glim} and are shown along with the results of other
experiments in Figure~\ref{fig:glresult}. As one may see from
Table~\ref{tab:glim} the number of photon candidates is
compatible with the background expectation $b$ obtained with the
proton Monte-Carlo.

\begin{table}[h]
\begin{center}
\begin{tabular}{|c|l|l|l|l|l|}
\hline
~   & \multicolumn{5}{c|}{$E_0$, eV} \\
~ & $~10^{18.0}$ & $~10^{18.5}$ & $~10^{19.0}~$ & $~10^{19.5}~$ & $~10^{20.0}~$ \\
\hline
$\gamma$ candidates & 1 & 0 & 0 & 0 & 1 \\
\hline
$b$ & 0.55 & 1.01 & 0.97 & 0.80 & 0.49 \\
\hline
$\bar n ~< $ & 5.14 & 3.09 & 3.09 & 3.09 & 5.14 \\
\hline
$A_{eff}$ & 77 & 255 & 852 & 2351 & 4055 \\
\hline
$F_\gamma$ $~<$ & 0.067 & 0.012 & 0.0036 & 0.0013 & 0.0013 \\
\hline
\end{tabular}
\end{center}
\caption{ \label{tab:glim} 95\% CL upper limits on the number
  of photons in the data set $\bar n_\gamma$ and on the photon flux
  $F_\gamma$ (km$^{-2}$yr$^{-1}$sr$^{-1}$). $b$ is an expected number
  of background photon candidates obtained with proton MC.}
\end{table}

Let us discuss the impact of the possible systematic
uncertainties. The result may be affected by the systematics of the
hadronic model as well as by the primary hadronic composition
different from proton assumed in simulations. Both the sources of
systematics act in a similar way by changing the background set used
for training and optimization of the cut. That is, the proton
Monte-Carlo is used to build a classifier and to define a criteria for
photon candidates. Let us note, however, that given the classifier we
are not using the proton Monte-Carlo in the final one-dimensional
analysis. In particular, the number of photon candidates expected from
the background $b$ (see Table~\ref{tab:glim}) is not used for
calculating the limit. The use of zero-background approximation makes
the result conservative with respect to the systematics of the
hadronic model and primary composition. On the other hand, the better
the data to Monte-Carlo agreement the better the sensitivity of the
method and the stronger the flux limits. We assume that the classifier
used in the Paper is pretty close to optimal based on the reasonable
agreement between data and proton Monte-Carlo, see
Figure~\ref{xi_dist}.

\section{Conclusion}
The use of the multiple observables within the multivariate classifier
and the wide range of zenith angles ($0^\circ<\theta<60^\circ$) along
with the nine years statistics allowed us to improve
significantly over the previous TA SD constraints on the diffuse
photon flux. The photon flux limits of the present {\it Paper} are the
most strict among the ones obtained in the Northern Hemisphere and
complement the limits set by Pierre Auger Observatory in the Southern
Hemisphere with the hybrid~\cite{Auger_hyblim} and SD
data~\cite{Auger_sdlim2}. The TA photon flux limits come close to the
values of the flux predicted in certain models of cosmogenic photons.

\section*{Acknowledgments}
{\small The Telescope Array experiment is supported by the Japan
  Society for the Promotion of Science(JSPS) through Grants-in-Aid for
  Priority Area 431, for Specially Promoted Research JP21000002, for
  Scientific Research (S) JP19104006, for Specially Promote Research
  JP15H05693, for Scientific Research (S) JP15H05741 and for Young
  Scientists (A) JPH26707011; by the joint research program of the
  Institute for Cosmic Ray Research (ICRR), The University of Tokyo;
  by the U.S. National Science Foundation awards PHY-0601915,
  PHY-1404495, PHY-1404502, and PHY-1607727; by the National Research
  Foundation of Korea (2017K1A4A3015188 ; 2016R1A2B4014967;
  2017R1A2A1A05071429, 2016R1A5A1013277);
by IISN project No. 4.4502.13, and Belgian Science Policy under IUAP
VII/37 (ULB). The development and application of the multivariate
analysis method is supported by the Russian Science Foundation grant
No. 17-72-20291 (INR).  The foundations of Dr. Ezekiel R. and Edna
Wattis Dumke, Willard L. Eccles, and George S. and Dolores Dore Eccles
all helped with generous donations. The State of Utah supported the
project through its Economic Development Board, and the University of
Utah through the Office of the Vice President for Research. The
experimental site became available through the cooperation of the Utah
School and Institutional Trust Lands Administration (SITLA),
U.S. Bureau of Land Management (BLM), and the U.S. Air Force.  We
appreciate the assistance of the State of Utah and Fillmore offices of
the BLM in crafting the Plan of Development for the site. Patrick Shea
assisted the collaboration with valuable advice on a variety of
topics. The people and the officials of Millard County, Utah have been
a source of steadfast and warm support for our work which we greatly
appreciate. We are indebted to the Millard County Road Department for
their efforts to maintain and clear the roads which get us to our
sites. We gratefully acknowledge the contribution from the technical
staffs of our home institutions. An allocation of computer time from
the Center for High Performance Computing at the University of Utah is
gratefully acknowledged. 
The cluster of the Theoretical Division of
INR RAS was used for the numerical part of the work. The lightning data used in this paper was obtained
from Vaisala, Inc. We appreciate Vaisala's academic research policy.
}

\end{document}